\documentstyle[preprint,prd,aps]{revtex}
\begin{document}
\draft
\title{Supersonic Deflagrations in Cosmological Phase Transitions}
\author{H.~Kurki-Suonio$^{a}$\thanks{Email: hkurkisu@pcu.helsinki.fi}
and M.~Laine$^{b}$\thanks{Email: mlaine@phcu.helsinki.fi}}
\address{$^a$Research Institute for Theoretical Physics and
$^b$Department of Physics, \\
P.O.~Box 9, FIN-00014 University of Helsinki, Finland}
\date{5 January 1995}
\maketitle

\vspace*{-6.5cm}
\hspace*{10.9cm} \mbox{Preprint HU-TFT-95-3}
\vspace*{6.0cm}

\begin{abstract}
The classification of the hydrodynamical growth mechanisms for the spherical
bubbles of the low-temperature phase in cosmological phase transitions is
completed by showing that the bubbles can grow as supersonic deflagrations.
Such deflagrations consist of a Jouguet deflagration, followed
by a rarefaction wave. Depending on the amount of supercooling,
the maximal velocity of supersonic deflagrations varies
between the sound and the light velocities. The solutions
faster than supersonic deflagrations are weak detonations.
\end{abstract}
\narrowtext

\section{Introduction}

When an exothermic process takes place in an ideal fluid,
knowing the hydrodynamics of growing bubbles becomes relevant.
Such situations include spherically symmetric chemical burning
in combustible gases~\cite{LL,CoFr}, and the growth of
bubbles of the low-temperature phase in cosmological first-order phase
transitions~\cite{St,GKKM,KS,KajKur,MilPan,EIKR,%
BonPan,HKLLM,KKT,Heckler,MilRez}.
The solutions of the energy-momentum
conservation equations across the `burning' surface can naturally
be divided into six hydrodynamical modes with different qualitative
features, called the weak, strong, and Jouguet deflagrations, and
the weak, strong, and Jouguet detonations. When stability
considerations, the microscopic mechanism of entropy production
at the burning surface, and the overall boundary conditions,
are taken into account, it turns out that some of the
hydrodynamical modes are actually forbidden. It is the purpose
of the present paper to complete the existing classification
of the allowed hydrodynamical modes in cosmological phase transitions,
and to present, to our knowledge, a new flow profile by which the
burning surface can be connected to the boundary conditions.

Our new results concern deflagrations,
which are the slower of the two main types of solutions.
Usually, it is assumed that the flow profile of a deflagration
bubble is such that the burning surface leaves the matter at rest.
Then weak deflagrations are subsonic and Jouguet deflagrations
expand at sound velocity. The strong deflagrations would be
supersonic, but it is known for chemical burning that such solutions
do not exist~\cite{LL,CoFr}. We find that the strong deflagrations
are forbidden in cosmological phase transitions, as well. However,
there is another type of a supersonic mode. Indeed, a Jouguet deflagration
can also leave behind it a rarefaction wave, in which case matter
is not at rest immediately behind the burning surface.
Such solutions are supersonic, and fill in the velocity gap between
the sound velocity and the slowest detonations. As far as we can see,
supersonic deflagrations should be possible both in chemical
burning, and in cosmological phase transitions, such as the QCD and
the electroweak phase transition.

The plan of the paper is the following.
In Sec.~\ref{defldet}, the properties of bubble growth
in relativistic ideal fluid are reviewed both for
deflagrations and detonations.
In Sec.~\ref{numerical}, we use a simple dynamical model
to study which hydrodynamical growth mechanisms can actually
be reached from an initial critical bubble configuration.
In particular, supersonic deflagrations are found to be possible.
The structure of spherical supersonic deflagration
bubbles is analyzed in detail
in Sec.~\ref{super}, for the bag equation of state.
The conclusions are in Sec.~\ref{concl}.

\section{Deflagrations and detonations}
\label{defldet}

Let us go into the rest frame of the expanding phase transition surface.
Imposing energy-momentum conservation across
the discontinuity gives the equations
\begin{eqnarray}
w_q\gamma_q^2v_q & = & w_h\gamma_h^2v_h \\
w_q\gamma_q^2v_q^2+p_q & = & w_h\gamma_h^2v_h^2+p_h
\,\, ,
\end{eqnarray}
where the subscripts $q$ and $h$ refer to the inflowing
high-temperature (`quark') phase and the out-flowing
low-temperature (`hadron') phase, respectively  ($w$ is the proper
enthalpy density).
We choose the signs so that the velocities
are positive, and denote by $F$ the
quantity $w_q\gamma_q^2v_q$. It then follows that
\begin{equation}
F=\frac{p_h-p_q}{v_q-v_h}>0
\,\, .
\end{equation}
This equation tells us that there are two qualitatively different kinds
of solutions. For {\em deflagrations}, the pressure decreases and
the velocity increases across the surface: $p_h<p_q$ and $v_h>v_q$.
For {\em detonations}, the opposite is true. For deflagrations,
the inflow velocity $v_q$ is subsonic, and for detonations, it is
supersonic~\cite{LL,La}. Note that for deflagrations, the sign of
the pressure difference is opposite to what it would have been, had
the temperature not changed across the phase transition surface.

A further division of the different hydrodynamical
processes can be made according to the outflow velocity~$v_h$.
Both for deflagrations and detonations,
energy-momentum conservation allows
$v_h$ to be on either side of, or equal to, the sound velocity~$c_s$.
If $v_h$ equals $c_s$, the process is called a {\em Jouguet process}.
If $v_h$ is on the same side of $c_s$
as $v_q$, the process is called {\em weak}; otherwise
it is called {\em strong}. For instance, for weak deflagrations, $v_h<c_s$.

To construct a complete bubble solution, the discontinuity at the phase
transition surface has to be matched with possible shock discontinuities
and continuous flow profiles, to satisfy the boundary conditions.
A deflagration solution can be constructed from a
supersonic shock front preceding the phase transition
surface and setting the cosmic matter moving, followed by
a continuous flow profile, and terminated by the phase transition
front bringing the matter back at rest. A detonation
solution can be constructed from a weak or Jouguet detonation front
moving supersonically into matter at rest, and
followed by a continuous rarefaction wave bringing
the matter again at rest. With a strong detonation,
the boundary conditions of a spherically expanding
bubble cannot be satisfied~\cite{LL,St}.
Flow profiles of deflagration bubbles are
given in Ref.~\cite{KS}, and flow profiles of detonation
bubbles are illustrated, e.g., in Fig.~2 of Ref.~\cite{La}.
These simplest complete flow profiles are dubbed,
according to the nature of the phase transition front, the weak,
Jouguet and strong deflagrations, and the weak and Jouguet
detonations. In the next Section, it is seen that
the presented solutions do not quite exhaust the flow profiles
relevant for cosmological phase transitions.
In the remainder of this Section, on the other hand,
it is argued that some of the solutions
will not be realized in nature.

First, strong deflagrations are expected to be forbidden.
For chemical combustion, there are at least two ways of seeing this.
Assuming that the burning front can be described as
a changing mixture of the burnt and the unburnt matter,
one can see that strong deflagrations
would include regions of negative entropy production~\cite{CoFr,La},
and they are therefore impossible.
This proof is not valid for cosmological phase transitions,
since the assumption of a changing mixture would lead to a vanishing
surface tension at $T_c$. There is another way of excluding
the strong deflagrations, coming from the observation that
even if such a solution could momentarily be forced to
exist, it would be unstable and tend to split into
other solutions~\cite{LL}. This argument should hold also
for cosmological phase transitions. Hence we expect to see
only weak and Jouguet deflagrations. These
expand at most at sound velocity.

Second, consider weak detonations.
To study them, one must know something about the microscopic
mechanism of phase separation.
For chemical combustion~\cite{LL,CoFr},
the rate of burning rises dramatically
with temperature, and the velocity of the burning
front is determined by how fast heat can be supplied
to yet unburnt regions. As a consequence,
the microscopic structure of a detonation front
is such that a shock front heats the matter
so much that rapid burning can begin, and a layer where the burning
takes place follows just behind the shock
front\footnote{This assumption was also made by Steinhardt in Ref.~\cite{St}.
In Ref.~\cite{St} weak detonations were excluded from consideration
by the statement
``The lower point $b$ at which the chord $ad$ intersects the detonation
adiabat cannot be reached if the reaction is exothermic'', which assumes
the detonation is initiated by a shock.
Ref.~\cite{St} is often
quoted as a proof that Jouguet detonations are the only kinds of
detonations that occur, but this is a result of relativistic combustion
theory, which does not apply to cosmological phase transitions.}.
For weak detonations, the layer behind the shock front
is a strong deflagration. Hence, the impossibility of strong deflagrations
implies the impossibility of weak detonations.
For cosmological phase transitions, the structure of
the burning front is different.
A phase transition is more apt to start at a {\em low} temperature,
when supercooling is larger. In particular, a point in space
outside the original critical bubble does not
wait for a rise in temperature to start transforming
into the new phase, but the transformation
is caused by interactions with the new phase
in the detonation front.
Hence, the structure of a detonation front need not be a shock front
followed by a deflagration, and weak detonations
cannot be excluded in cosmological phase transitions~\cite{La}.

Let us note that there exist processes in classical
hydrodynamics which have the same classification
of the different hydrodynamical growth mechanisms as
cosmological phase transitions, unlike chemical burning.
Such are {\em condensation discontinuities}~\cite[\S132]{LL}.
An example of a condensation discontinuity is the condensation of
supersatured water vapour into small water droplets, i.e., fog.
This is a phase transition, though with a finite chemical potential.
Indeed, only strong deflagrations and detonations are excluded
in Ref.~\cite{LL}, and the other processes are accepted.

\section{Numerical results}
\label{numerical}

To be able to check the presented arguments about the possibility
of the different kinds of solutions, one would like to have
a way of taking automatically into account energy-momentum
conservation, boundary conditions, and the microscopic
structure of the phase transition surface.
One might also wish to follow the real-time history of
the growing bubbles, to see which stationary final states can
actually be reached from reasonable initial conditions.
These problems can be solved by imposing an equation
of motion for some continuously changing
order parameter field $\varphi(t,{\bf x})$, and by then solving
the equation. The equation could, in principle, be derived
from non-equilibrium field theory.

In Ref.~\cite{IKKL}, the simple phenomenological form
\begin{equation}
\partial^{\mu}\partial_{\mu}\varphi+\frac{\partial V}{\partial\varphi}
+{\eta}u^{\mu}\partial_{\mu}\varphi=0
\label{Langevin}
\end{equation}
of the equation for the $\varphi$-field was proposed
(in Ref.~\cite{IKKL} the notation $1/\Gamma$ was used for $\eta$).
This is the simplest relativistic
generalization of the Langevin equation without the noise term.
The parameter $\eta$ fixes the one degree of freedom
left free by energy-momentum conservation equations.
{}From eq.~(\ref{Langevin}) and energy-momentum
conservation it follows that the entropy density~$s$ satisfies
\begin{equation}
T\partial_{\mu}(su^{\mu})={\eta}(u^{\mu}\partial_{\mu}\varphi)^2
\,\, ,
\end{equation}
so that $\eta$ can be interpreted as a parametrization of the entropy
production at the phase transition surface. Below, $\eta$ is treated as a
free parameter, although for the EW phase transition, its value could be
estimated~\cite{IKKL} by comparing the analytic formula for the deflagration
front velocity in the limit of small supercooling with microscopic
calculations~\cite{DHLLL,LMT,Kh} of the same quantity.

In Ref.~\cite{IKKL}, eq.~(\ref{Langevin}) and the energy-momentum conservation
equations were solved numerically for the planar symmetric case,
corresponding to a 1+1 dimensional spacetime.
Note that the dimension of space~$d$ does not affect the
results qualitatively, since
the phase transition front looks locally planar, and the
qualitative features of the continuous flow
profiles outside the phase transition front do not
depend on~$d$~\cite{KS}.
There are two methods of solving eq.~(\ref{Langevin}).
Either one can take some initial
bubble, and integrate all the partial differential equations forward
in time (the {\em dynamical} code),
or one can search only for stationary final states (the {\em static} code).

\begin{figure}
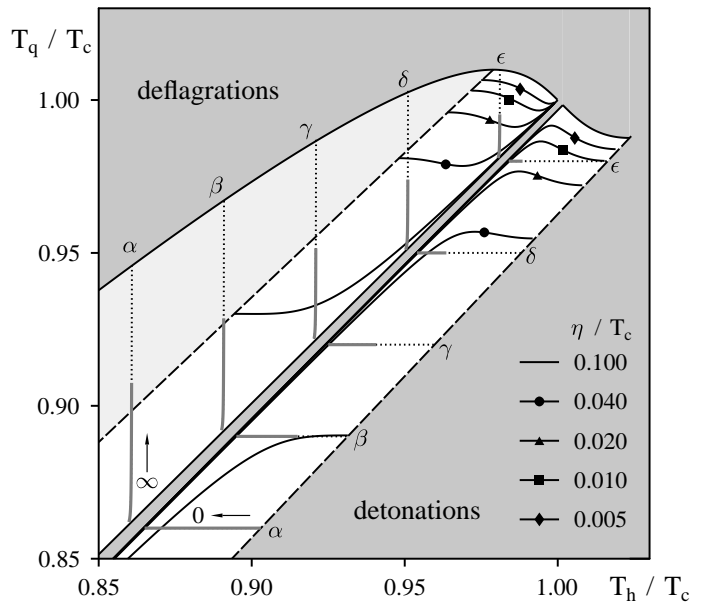

\caption[a]{\protect
The solutions of the model of Ref.~\cite{IKKL},
in the case of planar symmetry.
In 1+3 dimensions, the solid lines remain the same,
but the exact functional form of the dotted lines changes
a little. For detonations, the
$x$-axis is the temperature just behind the
phase transition surface; for deflagrations,
it is the temperature in the center of the expanding
bubble. The $y$-axis is the
temperature just in front of the phase transition surface.
The region forbidden kinematically, by the non-negativity
of entropy production, or by boundary conditions, has been
covered with the darker shade, assuming the types of solutions
discussed in Sec.~\ref{defldet}. The two allowed regions
represent the deflagrations and the weak detonations.
Dashed lines indicate the Jouguet processes, and
strong deflagrations are covered with
the lighter shade. The meaning of the different curves
has been explained in the text. The nucleation temperatures
$T_f/T_c$ corresponding to $\alpha$, $\beta$, $\gamma$, $\delta$ and
$\epsilon$ are $0.86$, $0.89$, $0.92$, $0.95$ and $0.98$.
The (QCD-type) parameters used in this
figure are $L=0.1T_c^4$, $\sigma=0.1T_c^3$
and $l_c=6T_c^{-1}$. Given the expansion rate of the Universe,
the actual nucleation temperature $T_f/T_c$ is then about
$0.86$ (Fig.~1 in Ref.~\cite{IKKL2}).}
\label{TfGamma}
\end{figure}

The general structure of the stationary
final states is shown in Fig.~\ref{TfGamma}.
In this figure, three kinds of solutions are displayed.
First, the dotted lines ($\alpha,\ldots$) each represent a one-parameter
family of solutions, with a given initial supercooling $T_f$,
which satisfy the energy-momentum conservation equations
and the boundary conditions, alone.
In drawing these curves, it was assumed that the solutions
are of the type discussed in Sec.~\ref{defldet},
Second, the solid lines represent the solutions which
satisfy energy-momentum conservation and eq.~(\ref{Langevin})
across the phase transition surface, with a given $\eta$,
but for which the global
boundary conditions have not been imposed. For given~$T_f$
and~$\eta$, the possible stationary final
states are then found from the intersections
of these two kinds of curves.
Third, the gray lines indicate those stationary final states which
are actually realized by the dynamical code,
starting from a reasonable initial state,
for given $T_f$, as a function of $\eta$. As illustrated
with the arrows for the curve~$\alpha$, the solution is
a weak deflagration in the limit $\eta\to\infty$, moves
towards strong deflagrations as $\eta$ is decreased,
and changes to a detonation at some point.

A few interesting things can be seen in Fig.~\ref{TfGamma}.
First, one should note the general feature
that weak detonations exist, as anticipated in Sec.~\ref{defldet}.
Second, the dotted lines have a very large gradient for
deflagrations. This indicates that the temperature $T_q$
in front of the phase transition surface may
be much higher than $T_f$
due to the heating caused
by the shock front, but the temperature $T_h$ behind
the phase transition surface is very near $T_f$.
For detonations, similarly, it turns out that
the temperature in the center of the
expanding bubble differs much less from $T_f$ than the temperature
$T_h$ just behind the detonation front.
Third, it is seen that for some $T_f$ and
$\eta$, there are more than one possible
stationary final states. For instance,
for $\eta =0.010T_c$ and $T_f=0.98T_c$,
there are three possible final states.
The dynamical code selects one of these\footnote{Figs.~19 and 20 of
Ref.~\cite{IKKL} represent one solution of the static code.
The solution of the dynamical code actually changes into a weak detonation at
around $\eta=0.03T_c$, with $v_{\rm det}\approx 0.7$.}.
The solutions which are not realized have a sharp peak in temperature
expanding at roughly the speed of sound. If one gives the
unreached weak deflagration as an initial condition to the dynamical code,
it remains as such, indicating that this solution may be reachable
from other initial conditions. However, the unreached weak detonation
transforms to either of the other solutions, suggesting an instability.

\begin{figure}
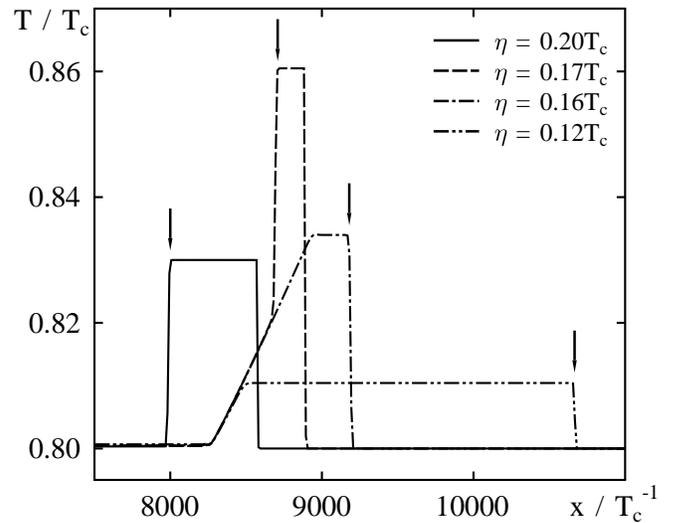

\caption[a]{\protect
Temperature profiles at $t=14400T_c^{-1}$ for different
values of $\eta$. All the profiles are moving to the right.
The arrows indicate the location of the phase transition front.
For $\eta=0.20T_c$, the solution is a weak deflagration;
for $\eta=0.17T_c$, it is a Jouguet deflagration followed
by a rarefaction wave;
for $\eta=0.16T_c$ and
$\eta=0.12T_c$, the solution is a weak detonation. The
velocity of the phase transition front changes
from $0.56$ at $\eta=0.20T_c$ to
$0.74$ at $\eta=0.12T_c$.}
\label{Gprofiles}
\end{figure}

Let us now turn to the existence of supersonic deflagrations.
{}From Fig.~\ref{TfGamma}, it is seen that strong deflagrations
have not been found with the static code. Technically, this is
due to the fact that no matter how well one tries to guess the
correct solution, the guess does not relax to a strong deflagration
but rather changes considerably to form a weak detonation,
with the same $T_h$. This is in accordance with strong deflagrations
being unstable, as argued in the previous Section.
On the other hand, with the dynamical code, we have found
supersonic solutions which lie very near the dotted lines
representing strong deflagrations (see curve $\alpha$).
A careful inspection reveals, however, that in the region
covered by the lighter shade, the gray curve is not
exactly on the dotted curve.  By drawing a figure
of the temperature profiles, the reason becomes apparent.
In Fig.~\ref{Gprofiles}, temperature profiles
are shown for the supercooling $T_f=0.80T_c$. The supersonic
deflagration solution~($\eta=0.17T_c$)
is seen to be comprised of a shock front,
a phase transition front, and a rarefaction wave
in the low-temperature phase. If the temperature
just behind the phase transition front is taken as the $x$-coordinate
in Fig.~\ref{TfGamma} instead of the temperature in the center
of the bubble, the solution lies on the dashed curve
corresponding to a Jouguet deflagration. Furthermore,
if one gives a strong deflagration as an initial condition to
the dynamical code, the solution is unstable, and
transforms into a Jouguet deflagration followed by a rarefaction
wave. In the next Section, these solutions are analyzed in more detail
in 1+3 dimensions.

\section{Spherical solutions}
\label{super}

   For an overview of the different hydrodynamical solutions for
spherical phase transition bubbles, we employ a simpler equation of
state, the bag EOS
\begin{eqnarray}
   p_h & = & {1\over3}\epsilon_h, \\
   p_q & = & {1\over3}(\epsilon_q-4B),
\end{eqnarray}
where $B$ is the bag constant ($\epsilon$ is the proper energy density).
We assume a spherically symmetric
similarity solution, {\it i.e.}, it can be described in terms of the
single coordinate $\xi \equiv r/t$.  In contrast to the previous Section,
we are here not interested in how the bubble relaxes to the similarity
solution from some initial condition.  We also give up modelling the
microscopic structure of the phase transition front, and treat it as a
discontinuity.  Therefore we do not have a parameter $\eta$ coming
from microphysics to select among the different solutions hydrodynamically
allowed.  We just parametrize the solutions by  the velocity of the
phase transition front $\xi_{\rm p.t.}$ ($\xi_{\rm defl}$ or $\xi_{\rm det}$).
Then we get almost analytical ({\it i.e.}, involving ordinary
differential equations) solutions for the bubble structure, instead of
having to resort to numerical hydrodynamics.

   The continuous parts of the flow satisfy the equations~\cite{KS}
\widetext
\begin{equation}
   \bigl[v^2(3-\xi^2)-4v\xi+3\xi^2-1\bigr]{dv\over d\xi}
   = 2{v\over\xi}(1-v^2)(1-\xi v)
\end{equation}
\narrowtext
and
\begin{equation}
   {1\over\tilde\epsilon}{d\tilde\epsilon\over d\xi}
   = 4{\xi-v\over1-\xi v}{1\over1-v^2}{dv\over d\xi},
\end{equation}
where $\tilde\epsilon = \epsilon$ in the $h$ phase and
$\tilde\epsilon = \epsilon - B$ in the $q$ phase.

\begin{figure}
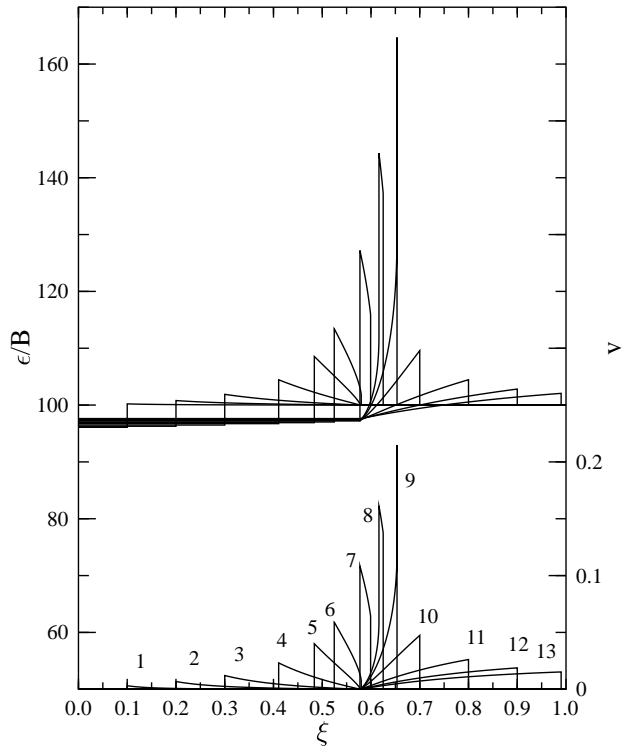

\caption[a]{\protect
   Energy-density (upper set) and flow velocity (lower set) profiles for
spherical phase transition bubbles.  We show 13 profiles, all for the
same initial energy density $\epsilon_f = 100 B$, but with different
phase transition front velocities.  The six slowest (on the left) are weak
deflagrations.  The seventh is a Jouguet deflagration bubble, with
$\xi_{\rm defl} = c_s$.  The eighth is a supersonic deflagration bubble,
which has a Jouguet deflagration front preceded by a shock and followed by
a rarefaction wave.  The ninth is a supersonic deflagration bubble
almost at the limit where it becomes a Jouguet detonation.  The shocked
region is now so thin that it appears as a mere vertical line in the
figure.  The last four (on the right) are weak detonations.}
\label{sudprset}
\end{figure}

   In Fig.~\ref{sudprset} we show a sequence of profiles for phase
transition bubbles, with the same initial energy density $\epsilon_f$,
but for different $\xi_{\rm p.t.}$.

   If the phase transition front is
subsonic, $\xi_{\rm p.t.} = \xi_{\rm defl}
\leq c_s$, we have a deflagration bubble
of the type described in \cite{KS}.
The phase transition front is preceded
by a shock at $\xi_{\rm sh} > c_s$, the
matter is at rest at $\xi > \xi_{\rm sh}$ and at $\xi < \xi_{\rm defl}$,
and the shocked fluid in between has a continuous flow profile, with the
(outward) flow velocity and energy density increasing with decreasing $\xi$.
We denote by $\xi_{\rm sh,J}$ the shock velocity of the Jouguet
deflagration bubble with $\xi_{\rm p.t.} = c_s$.
The slowest deflagrations have extremely weak shocks at $\xi_{\rm sh}
\sim c_s$.

   The slowest detonation is the Jouguet detonation, with
\begin{equation}
   \xi_{\rm det,J} = {1 + 3\sqrt{B \over 2\epsilon_f + B}
      \over 1 + \sqrt{B \over 2\epsilon_f + B}} c_s.
\end{equation}
Solutions with $\xi_{\rm p.t.} > \xi_{\rm det,J}$ are weak detonations.

   The velocity range between $\xi_{\rm p.t.} = c_s$ and
 $\xi_{\rm p.t.} = \xi_{\rm det,J}$ is covered by the supersonic
deflagrations.  As $\xi_{\rm p.t.}$ is increased past $c_s$, a
rarefaction wave appears in the region between $\xi = c_s$
and $\xi = \xi_{\rm p.t.}$.  The phase transition remains a Jouguet
deflagration, and is preceded by a shock, with $\xi_{\rm p.t.} <
\xi_{\rm sh} < \xi_{\rm det,J}$.  As $\xi_{\rm p.t.}$ is increased,
$\xi_{\rm sh}$ grows also, but not as rapidly, and thus the region of
shocked fluid becomes narrower.  The two velocities
$\xi_{\rm p.t.}$ and $\xi_{\rm sh}$
become equal at  $\xi_{\rm p.t.} = \xi_{\rm sh} = \xi_{\rm det,J}$,
the region of shocked fluid disappears and the solution becomes a Jouguet
detonation.

   In the limit where $\xi_{\rm p.t.}$ approaches $\xi_{\rm det,J}$ from
below, the shocked region becomes infinitesimally thin.  This resembles
the microscopic structure of a detonation front in chemical combustion,
where heating by a shock initiates the transition.
However, the weak detonations ($\xi_{\rm p.t.} > \xi_{\rm det,J}$) here
do not have such structure.

\begin{figure}
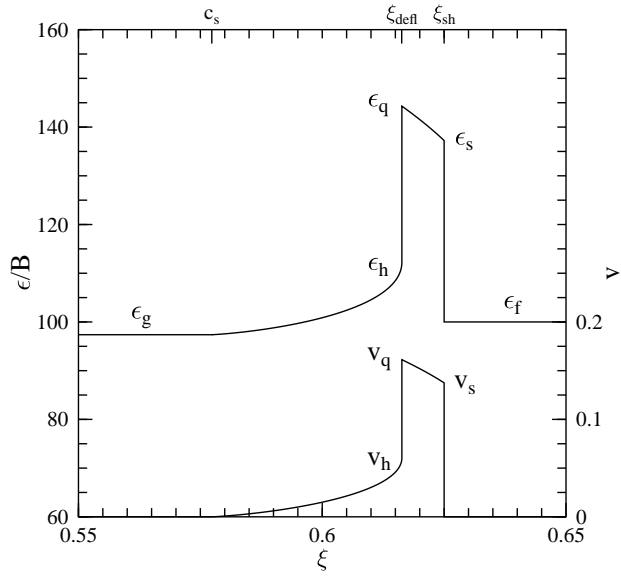

\caption[a]{\protect
   A supersonic deflagration bubble.  The upper curve is the energy density
profile and the lower curve the flow velocity profile.  This figure
is for a supersonic
spherical deflagration bubble with an initial energy density
$\epsilon_f = 100 B$ and shock velocity $\xi_{\rm sh} = 0.625$.
The velocity of the phase transition front is $\xi_{\rm defl} = 0.6163$.}
\label{sudpr}
\end{figure}

\begin{figure}
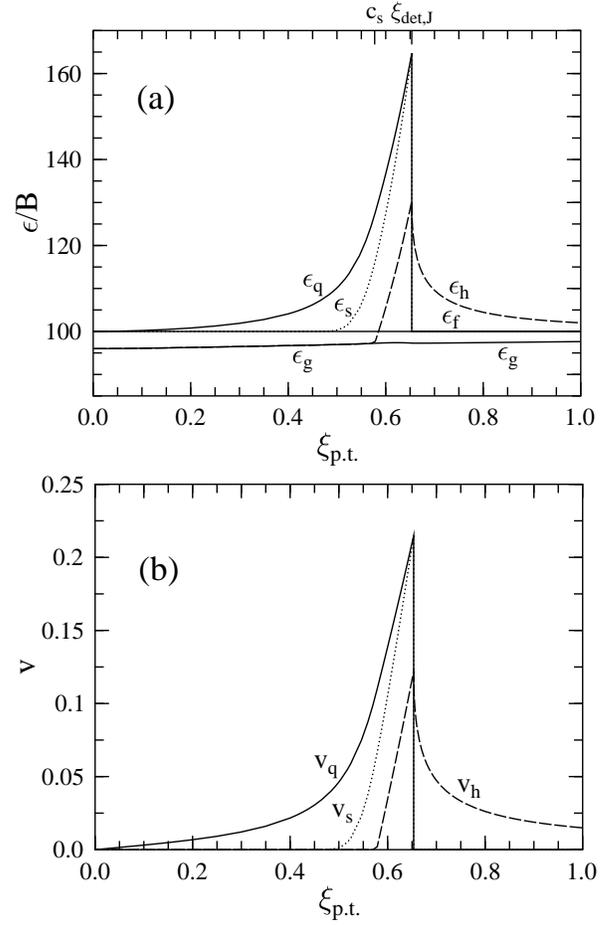

\caption[a]{\protect
   The energy densities (a) and flow velocities (b) at special points of
the bubble profile (see Fig.~4).  These quantities are shown
as a function of the phase transition velocity  $\xi_{\rm p.t.}$ for a
fixed initial energy density $\epsilon_f = 100 B$.  The solution
switches from a weak deflagration bubble to a supersonic deflagration
bubble at $\xi_{\rm p.t.} = c_s = 0.57735$ and to a weak detonation
bubble at $\xi_{\rm p.t.} = \xi_{\rm det,J} = 0.6534$.}
\label{evgraph}
\end{figure}

\begin{figure}
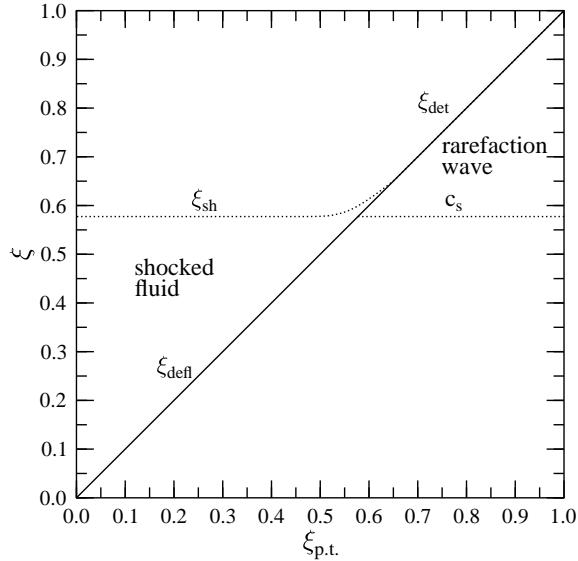

\caption[a]{\protect
   The special points of the bubble profile as a function of  $\xi_{\rm
p.t.}$ for $\epsilon_f = 100 B$.  A weak deflagration bubble has a region
of shocked fluid between $\xi = \xi_{\rm defl}$ and  $\xi = \xi_{\rm
sh}$.  A weak detonation bubble has a rarefaction wave extending from
$\xi = c_s$ to $\xi = \xi_{\rm det}$.  A supersonic deflagration bubble
has both features.}
\label{xgraph}
\end{figure}

\begin{figure}
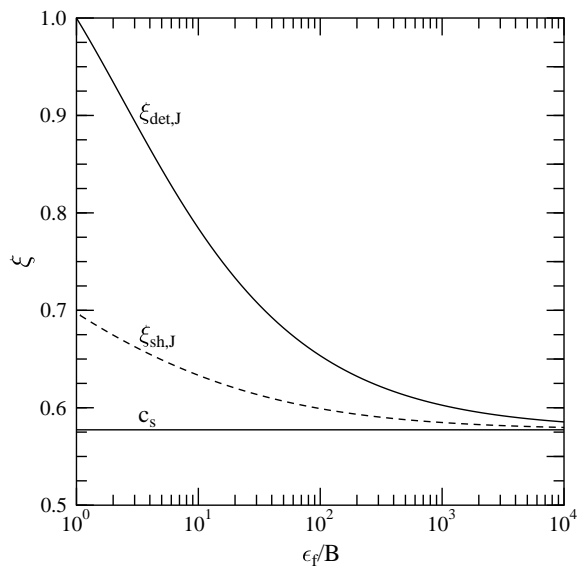

\caption[a]{\protect
   The velocity $\xi_{\rm det,J}$ of the phase transition front in the Jouguet
detonation bubble and the shock velocity $\xi_{\rm sh,J}$ in the Jouguet
deflagration bubble as a function of the initial
energy density~$\epsilon_f$.
As the velocity of a weak deflagration is increased from
$0$ to $c_s$, the
shock velocity increases from $c_s$ to $\xi_{\rm sh,J}$, at which point
the solution becomes a supersonic deflagration bubble.  As the deflagration
velocity is further increased from $c_s$ to $\xi_{\rm det,J}$ the shock
velocity increases from $\xi_{\rm sh,J}$ to $\xi_{\rm det,J}$ at which
point the phase transition front catches up with the shock and the solution
becomes a detonation bubble.}
\label{limcases}
\end{figure}

   In Fig.~\ref{sudpr} we show the structure of a supersonic deflagration
bubble.  Figs.~\ref{evgraph} and \ref{xgraph} show
how the different quantities in the
bubble change as a function of $\xi_{\rm p.t.}$.  Fig.~\ref{limcases}
shows the limiting velocities $\xi_{\rm sh,J}$ and $\xi_{\rm det,J}$ as
a function of $\epsilon_f / B$.

   Let us note that in this Section we have studied pure hydrodynamics,
whereas one should actually also insist on entropy production being
non-negative at the phase transition front~\cite{GKKM}.
This requires knowledge
of the dimensionless parameter $r = a_q/a_h$ in the bag EOS ($\epsilon_h
= 3 a_h T^4$, $\epsilon_q = 3 a_q T^4 + B$).  The parameter $r$ fixes the
energy density $\epsilon/B = (4r-1)/(r-1)$ corresponding to the critical
temperature $T_c$.  The entropy condition restricts the availability of the
different hydrodynamical modes near $T_c$, and
for a sufficiently small supercooling, only weak deflagrations
remain allowed~\cite[Sec.~IV]{IKKL2}.
However, considering any $\epsilon_f$, for a sufficiently small $r$
this value of $\epsilon_f$ will correspond to a large supercooling, and
all the hydrodynamical modes are allowed.  Therefore the entropy
condition does not change our conclusions.

\section{Conclusions}
\label{concl}

   We have discovered a new kind of a bubble solution for cosmological
phase transitions.  The phase transition front of such a bubble is a Jouguet
deflagration, is preceded by a shock, and followed by a rarefaction
wave.  The phase transition front moves supersonically with respect to
the matter at rest.  These supersonic deflagration bubbles fill the
velocity gap between weak deflagrations, which are subsonic, and weak
detonations.  The two limiting cases of the supersonic deflagration
bubble are the `ordinary' Jouguet deflagration bubble ({\it i.e.},
without a rarefaction wave) and the Jouguet detonation bubble.

   It has been known that strong deflagrations would not occur since
they are unstable to 1-dimensional perturbations ({\it i.e.}, in the
radial direction).  By following a strong deflagration bubble with a numerical
hydrodynamics code we have seen how such a perturbation transforms
the strong deflagration into a Jouguet deflagration by developing a
rarefaction wave.  This gives rise to the supersonic deflagration bubble
we have described.

   By modelling the microscopic mechanism of entropy production at the
phase transition front by a simple dynamical model,
we found that the solutions
`near' Jouguet detonations [{\it e.g.}, profiles 7--9 in
Fig.~\ref{sudprset}] were in many cases not realized in a numerical
evolution, although they were allowed by the external conditions
(including the entropy condition).
The velocity $\xi_{\rm p.t.}$ of the phase transition front was found not to
increase continuously as the microscopic entropy-production parameter
$\eta$ was decreased, but the solution jumped from a weak deflagration
to a much faster weak detonation.  Thus the most `violent' solutions,
{\it i.e.}, the ones with large energy densities and flow velocities
occurring near the phase transition front, were skipped over.
This happened in the cases with a small supercooling, and we had to
go to large supercooling to see examples of supersonic deflagrations.
Examining Fig.~\ref{TfGamma}, one finds for small supercooling that,
for a given value of $T_f$ and $\eta$ there can be as many as three solutions
allowed, not just one.  One of these is a deflagration, and two are
detonations.  In the numerical evolution, the weaker, {\it i.e.}, the
faster, of the detonations was then realized.

   This suggests that in nature the more dramatic solutions, the
Jouguet detonation and solutions near it, are more difficult to
realize, especially for small supercooling.  This is in sharp contrast
to chemical burning, where the Jouguet detonations are the only possible
detonation solutions.

\section*{Acknowledgements}

We are grateful to J. Ignatius for discussions, and to the
Finnish Center for Scientific Computing for computational resources.

\end{document}